# Simultaneous Detection of Multiple Appliances from Smart-meter Measurements via Multi-Label Consistent Deep Dictionary Learning and Deep Transform Learning

Vanika Singhal, Jyoti Maggu, and Angshul Majumdar, *Senior Member, IEEE*

*Abstract*-- Currently there are several well-known approaches to non-intrusive appliance load monitoring – rule based, stochastic finite state machines, neural networks and sparse coding. Recently several studies have proposed a new approach based on multi-label classification. Different appliances are treated as separate classes, and the task is to identify the classes given the aggregate smart-meter reading. Prior studies in this area have used off-the-shelf algorithms like MLKNN and RAKEL to address this problem. In this work, we propose a deep learning based technique. There are hardly any studies in deep learning based multi-label classification; two new deep learning techniques to solve the said problem are fundamental contributions of this work. These are deep dictionary learning and deep transform learning. Thorough experimental results on benchmark datasets show marked improvement over existing studies.

*Index Terms*-- deep learning, energy disaggregation, multi-label classification, non-intrusive load monitoring.

## I. Introduction

TODAY there is a concerted effort towards sustainable Energy. On one hand research is carried out on alternate sources of energy (Solar, Hydel, Wind etc.) to make them commercially viable. On the other hand there is an effort to save energy. Non-intrusive load monitoring (NILM) belongs to the later. Technically the goal is to disaggregate the energy consumption of each appliance from the aggregate smart-meter data. In the broader perspective, this information is fed back to the consumer, so that he/she can make an informed choice about saving energy wherever possible. Since residential and commercial buildings account for 40% of the global energy consumption [1]; such smart load management is expected to significantly save power.

Appliance level load monitoring can be broadly categorized as, intrusive load monitoring and non-intrusive load monitoring (NILM) [2]. The former is expensive and cumbersome to implement and requires installation of sensors on each appliance; but intrusive load monitoring yields the gold standard. Ideally NILM should be able to disaggregate the load from only the smart-meter reading. However, this is a very difficult problem (especially at low sampling rates); in practice, a learning based paradigm is followed where the training stage is intrusive, but the testing / operation stage is not. The building is instrumented during the training stage gather data, from which machine learning models are learnt. During operation, the sensors are removed, and the learnt models used to predict the consumption of each device.

This aforesaid paradigm is not fully non-intrusive. The training stage is intrusive requiring deployment of multiple sensors. In recent times, a multi-label classification approach offers a fully non-intrusive alternative [3-6]. It does not require any instrumentation; it only requires the recording of the state-of-the-appliance, i.e. whether it is ON or OFF. During the training stage, given the smart-meter reading and the recorded states of the appliances, a machine learning model learns multi-label classification; here each appliance is treated as a label – since several appliances can be ON at the same time, it turns out to be a multi-label classification problem. During the operational stage, the learnt model is used to predict the state of the appliances given the smart-meter readings.

One might argue that such techniques do not estimate the actual energy consumption of the device. This can be addressed by multiplying the state of the appliance with the average power consumption of the device.

Prior studies have used off-the-shelf machine learning algorithms for multi-label classification [7-11], e.g. Multi Label K Nearest Neighbor (MLKNN), RAKEL (Random K-label sets). Our work is motivated by the success of deep learning. Recent studies in almost all areas of data analysis shows widespread insurgence of deep learning – mainly owing to its superior results over traditional (shallow) machine learning techniques. However, standard deep neural network models based on stacked autoencoder or deep belief networks are not capable of handling multi-label classification; this is because the output layer is trained by logistic regression or soft-max – both of which lead to single classes. Therefore such existing deep learning tools cannot be used for the current purpose.

In this work, we propose two new deep learning techniques

This work is supported by the DST IC-IMPACTS grant on Energy and Water Disaggregation for Non-Intrusive Load Monitoring in Buildings.

V. Singhal, J. Maggu and A. Majumdar are with Indraprastha Institute of Information Technology, Delhi 110020 India (e-mail: vanikas, jyotim, angshul@iiitd.ac.in).

for multi-label classification. Given the shortcomings of existing deep learning frameworks, we propose to build two new multi-label deep learning tools based on dictionary learning and transform learning. Experiments have been carried out on two benchmark NILM datasets. The results show significant improvement over existing techniques.

Rest of the paper is organized into several sections. The literature review on NILM and background of the proposed techniques is covered in the following section. Details of the proposed techniques is explained in the section 3. The experimental results are shown in section 4. The conclusions are drawn in section 5.

## II. LITERATURE REVIEW

In this section, we will discuss both non-intrusive load monitoring methods and the signal processing / machine learning background required for this work.

### A. Multi-label Classification for NILM

Multi-label classification algorithms have shown significant performance in identifying active appliances during a time period. In binary label classification one sample belongs to only one output class, whereas in multi-label problem a sample may belong to more than one output classes. The second scenario is appropriate for NILM since at a given point of time it is likely that many appliances are running. Note that simplifying this to a single label multi-class classification problem has been attempted before [12].

Multi-label classification algorithms can be divided into two categories: problem transformation and algorithm adaptation. Problem transformation methods transform a multi-label problem into multi-class classification problem. Examples of such methods are Label power-set (LP) [7] and binary relevance [13], which uses single label classifier like SVM as base classifiers. On the other hand, algorithm adaptation methods work by modifying the single label classification algorithms to deal with multi-label problems. E.g. MLKNN is a multi-label classification methods, it is derived from k-nearest neighbor algorithm [9].

In binary relevance, a separate binary classifier is trained for each label, and the results from all the classifiers are combined to produce the final output. The disadvantage of this method is that it does not consider any label dependency and fail to predict label combinations when some dependency exists. Label powerset (LP) trains a classifier for each pair of labels. It takes label correlation into account but become very complex when number of classes increase. RAKEL overcomes the disadvantage of label powerset method [7]. It breaks large set of labels into smaller label sets, and each of them is solved by applying LP method. MLKNN is an algorithm adaptation method derived from KNN algorithm [9].

In recent years NILM problem has been modeled as a multi-label classification problem. In [3], temporal sliding window technique is employed to extract features from the aggregated power data. Binary relevance, classifier chains and LP classifiers (SVM and decision tree as base classifiers) are trained using extracted features. The paper [5], uses delay embedding to extract features from the time series data and compare the results using multi-label classification algorithms. In [3, 5], the performance of RAKEL and MLKNN are compared on time series and Haar wavelet features extracted from the aggregated power data.

The primary shortcoming of all previous multi-label classification techniques is that they consider all combination of classes as a separate class. This leads to combinatorial growth in the number of classes. This in turn leads to the exponential growth in model parameters. With limited training volume, training so many parameters leads to over-fitting. This leads to poor results in real applications.

### B. Rule Based Techniques

If one has access to high frequency data, rule based methods offer a good solution. For example, in a typical household, periodic cycles throughout the day in power consumption may be related to the refrigerator. In the evenings, a different kind of cycle can be related to air conditioner (AC). In rule based techniques, instead of figuring out the rules manually traditional artificial intelligence techniques are used to learn them.

One popular approach in rule based systems are Decision Trees. Decision Tree uses greedy hunt's algorithm [4]. It evaluates impurities on each node and then best split among attributes is decided by Gini index; which results in the nodes with lowest value of impurity. The decision tree based classification is implicitly a binary classification problem. But, in NILM classification, which is a multiclass problem, decision tree based approach cannot be directly used [4, 14]. So, one vs rest strategy is applied where one class is considered as positive class and all other classes belong to negative class. In this way, multiple decision trees are built to solve the multiclass problem.

However, one must note that such rule based systems are only successful when the sampling frequency is very high. In practice, the sampling frequencies are very low, the smart-meter transmits the reading once every 10 or 15 minutes. At such temporal resolutions, the sharp edges required for rule based systems are flattened out and the rule based methods perform poor. Also such rule based systems are good for binary state (Fan, CFL, etc.) or multi-state (washer, dryer, AC etc.) appliances; they cannot handle continuously varying loads like printers or computers.

### C. Stochastic Finite State Models

Early studies in NILM [2] modeled appliances as finite state machines. Later on it was realized that stochastic finite state machines are more suitable owing to noise in the data. Since the state of the appliances vary dynamically, Hidden Markov Model (HMM) became an adequate tool for NILM classification when there is only one appliance. As the name suggests, HMM is based on the Markovian assumption – the current state of the appliance is dependent on the previous state. The HMM learns the state of the appliance given the observed readings.

Typical NILM scenario consists of multiple appliances,

hence factorial hidden Markov model (FHMM) is used. In FHMM [15, 16], a separate and independent HMM represents each appliance and complex information can be captured by combining outputs from all the HMMs. The product of expert is closely related to FHMM.

The issues that hinder the performance of rule based systems are also present in FHMM. They can be used for appliances where there is a marked difference in power consumption across the states, but they fail to model continuously varying appliances. Besides, HMM based techniques also rely on high frequency data that is impractical in most scenarios.

### D. Neural Networks

Traditional neural networks could not handle multi-label classification problems. This is largely because of the choice of the supervision penalties like logistic regression or softmax. They were only applicable for single label classification problems. However a recent study proposed a smart solution to the problem by learning one neural network for each load [17]. To the best of our understanding, the neural networks are run on the aggregated data, where each device specific neural network identifies if that device is turned ON or not.

There is only one published study [18] that uses stacked autoencoders for multi-label classification. They learn a map from the deepest layer of the encoders to a multi-label target. The idea has been first proposed in [19], but was used therein for single label classification problems; it was generalized in [18].

### E. Dictionary Learning

To overcome the issues with low-frequency sampling and continuously varying appliances, a recent class of methods based on dictionary learning have been proposed.

Since dictionary learning is directly pertinent to this work, we will discuss it in some detail. Kolter et al [20] introduced dictionary learning to solve disaggregation problems. The study assumed that there is training data collected over time, where the smart-meter logs only consumption from a single device only. This can be expressed as $X_i$ where $i$ is the index for an appliance, the columns of $X_i$ are the readings over a period of time. For each appliance ($i$) a basis ($D_i$) is learnt, such that the data ($X_i$) can be regenerated from the associated coefficients ($Z_i$)

$$X_i = D_i Z_i, \ i=1...N \qquad (1)$$

This is a typical dictionary learning problem with sparse coefficients. It can be solved via the following minimization:

$$\min_{D_i, Z_i} \|X_i - D_i Z_i\|_F^2 + \lambda \|Z_i\|_1 \qquad (2)$$

Learning the basis, constitutes the training phase. During actual operation, several appliances are likely to be in use simultaneously. Dictionary learning based techniques make the assumption that the aggregate reading by the smart-meter is a sum of the powers for individual appliances. Thus, if $X$ is the total power from N appliances (where the columns indicate smart-meter readings over the same period of time as in training) the aggregate power is modelled as:

$$X = \sum_i X_i = \sum_i D_i Z_i \qquad (3)$$

Given this model, it is possible to find out the loading coefficients of each device by solving the following sparse recovery problem,

$$\min_{Z_1,...,Z_N} \left\| X - [D_1 | ... | D_N] \begin{bmatrix} Z_1 \\ ... \\ Z_N \end{bmatrix} \right\|_F^2 + \lambda \left\| \begin{matrix} Z_1 \\ ... \\ Z_N \end{matrix} \right\|_1 \qquad (4)$$

Here a positivity constraint on the loading coefficients is enforced as well. This is a convex problem since the basis are fixed. Once the loading coefficients are estimated, one can easily compute the power consumption from individual devices.

$$\hat{X}_i = D_i Z_i, \ i=1...N \qquad (5)$$

This was proposed as the initial technique in [20]. They proposed other formulations where discrimination was introduced. However, such added penalties did not improve the overall results significantly.

Dictionary Learning based techniques in disaggregation has been gaining popularity ever since. In [21] a dynamic model is incorporated into the dictionaries. The most recent work in this topic is deep sparse coding for energy disaggregation [22]; they proposed a deep sparse coding framework by learning multiple levels of dictionaries for each device.

### F. Transform Learning

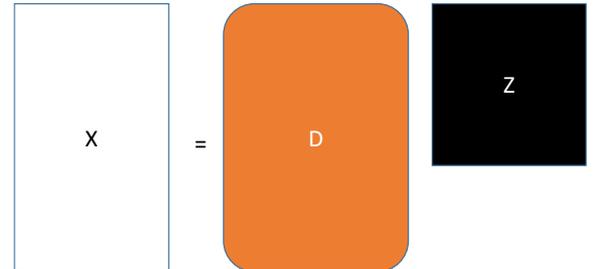

Fig. 1. Schematic Diagram for Dictionary Learning

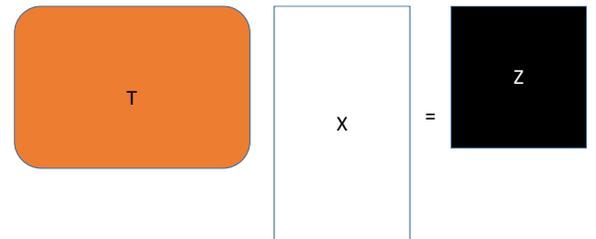

Fig. 2. Schematic Diagram for Transform Learning

Dictionary learning is a synthesis formulation. It learns a dictionary ($D$) such that it can synthesize the data ($X$) from the learnt coefficients ($Z$) (see Fig. 1). Mathematically it is expressed as (1),

$$X = DZ$$

Transform Learning is the analysis equivalent of dictionary learning. It learns an analysis dictionary / transform ($T$) such that it operates on the data ($X$) to generate the coefficients ($Z$) (see Fig. 2). Mathematically this is represented as,

$$TX = Z \qquad (7)$$

(Synthesis) dictionary learning is very popular in signal processing and machine learning. It has hundreds of papers in each area. Transform learning on the other hand is relatively new. There are hardly any papers on this topic outside signal processing. Therefore, we discuss it is slightly greater detail.

One may be enticed to solve the transform learning problem by formulating,

$$\min_{T,Z} \|TX - Z\|_F^2 + \mu \|Z\|_0 \quad (8)$$

Unfortunately, such a formulation would lead to degenerate solutions; it is easy to verify the trivial solution $T=0$ and $Z=0$. In order to ameliorate this the following formulation was proposed in [23] –

$$\min_{T,Z} \|TX - Z\|_F^2 + \lambda \left( \varepsilon \|T\|_F^2 - \log \det T \right) + \mu \|Z\|_0 \quad (9)$$

The factor $-\log \det T$ imposes a full rank on the learned transform; this prevents the degenerate solution. The additional penalty $\|T\|_F^2$ is to balance scale; without this $-\log \det T$ can keep on increasing producing degenerate results in the other extreme.

Note that the sparsity constraint on the coefficients is not mandatory for machine learning problems. It is useful for solving inverse problems in signal processing.

In [23], [24], an alternating minimization approach was proposed to solve the transform learning problem (9).

$$Z \leftarrow \min_Z \|TX - Z\|_F^2 + \mu \|Z\|_0 \quad (10a)$$

$$T \leftarrow \min_T \|TX - Z\|_F^2 + \varepsilon \left( \|T\|_F^2 - \log \det T \right) \quad (10b)$$

Updating the coefficients (10a) is straightforward. It can be updated via one step of Hard Thresholding [25]. This is expressed as,

$$Z \leftarrow (abs(TX) \geq \mu) \odot \quad (11)$$

Here $\odot$ indicates element-wise product.

For updating the transform, one can notice that the gradients for different terms in (10b) are easy to compute. After ignoring the constants, the gradients are given by –

$$\nabla \|TX - Z\|_F^2 = X^T (TX - Z)$$

$$\nabla \|T\|_F^2 = T$$

$$\nabla \log \det T = T^{-T}$$

In the initial paper on transform learning [23], a non-linear conjugate gradient based technique was proposed to solve the transform update. In the second paper [24], with some linear algebraic tricks they were able to show that a closed form update exists for the transform.

$$XX^T + \lambda \varepsilon I = LL^T \quad (12a)$$

$$L^{-1}YX^T = USV^T \quad (12b)$$

$$T = 0.5R \left( S + (S^2 + 2\lambda I)^{1/2} \right) Q^T L^{-1} \quad (12c)$$

The first step is to compute the Cholesky decomposition; the decomposition exists since $XX^T + \lambda \varepsilon I$ is symmetric positive definite. The next step is to compute the full SVD. The final step is the update step. One must notice that $L^{-1}$ is easy to compute since it is a lower triangular matrix.

The proof for convergence of such an update algorithm can be found in [26]. It was found that the transform learning was robust to initialization.

### III. PROPOSED METHODOLOGY

Posing NILM as a multi-label classification problem is relatively new. As mentioned before, prior studies used existing algorithms and applied it on the NILM datasets to publish papers. Our work is motivated by the success of deep learning in almost all areas of data science and artificial intelligence.

However standard deep neural networks are either unsupervised or can only handle multi-class problems; they are not geared for multi-label classification. Therefore, in this work we propose two new deep learning approaches for multi-class multi-label classification.

The first approach is based on the deep dictionary learning paradigm [27]. It has been introduced recently. It performs better than other deep learning techniques like stacked autoencoder and deep belief network on a variety of problems [28-30]. However deep dictionary learning has been an unsupervised learning tool so far. This will be the first work on supervised deep dictionary learning.

#### A. Multi-Label Consistent Deep Dictionary Learning

Since deep dictionary learning is a new approach, we will briefly review it before proposing out formulation. In standard (shallow) dictionary learning, one level of dictionary is learnt to represent the training data (1). We repeat it for the sake of convenience.

$$X = D_1 Z$$

In deep dictionary learning, multiple levels of dictionaries are used as basis for representing the data. In (13) we show it for three levels.

$$X = D_1 \varphi \left( D_2 \varphi (D_3 Z) \right) \quad (13)$$

Here $D_1$, $D_2$ and $D_3$ are the three level dictionaries. The activation function $\varphi$ assures that the three levels are not collapsible into one. There have been recent studies on deep matrix factorization [31], [32]; these methods do not have the activation function. Deep dictionary learning is a generalization of deep matrix factorization for arbitrary activation functions.

Usually in deep dictionary learning, the dictionaries are learnt by solving the following optimization problems –

$$\min_{D_1, D_2, D_3, Z} \|X - D_1 \varphi (D_2 \varphi (D_3 Z))\|_F^2 \quad (14)$$

This is an unsupervised formulation. It does not use any class information. One can imagine (14) in terms of a neural network like interpretation, where the coefficients from the shallower levels feeds into the deeper levels. $Z$ is the final level of coefficients. Note that deep dictionary learning is akin to a feed backward neural network.

In this work, we propose a supervised multi-label variant of deep dictionary learning. Taking cues from label consistent dictionary learning [33], we propose a multi-label consistency term. Basically, we learn a linear map such that the coefficients from the final level maps to the multi-label targets. This is expressed as,

$$\min_{D_1,D_2,D_3,Z,M} \|X - D_1\varphi(D_2\varphi(D_3Z))\|_F^2 + \lambda\|T - MZ\|_F^2 \quad (15)$$

Here $T$ are the targets. Each target has the same length as the number of appliance; the appliances are in an order. If an appliance is ON, the corresponding value is 1 or else it is 0. The map $M$, projects the coefficients $Z$ to the multi-label target labels $T$.

Solving (15) is not trivial. In the first work on dictionary learning, it has been solved greedily one layer at a time. This is sub-optimal; the shallower layers influence the deeper layers but not the other way around. In an optimal solution all the variables should be influencing each other. The majorization minimization (MM) techniques used by Singhal and Majumdar [28] cannot be used either owing to the supervision term. Besides MM converges very slowly. In this work we solve it using the Augmented Lagrangian (AL) approach [34].

We substitute $Z_2 = \varphi(D_3Z)$, $Z_1 = \varphi(D_2Z_2)$. This leads to the following AL formulation.

$$\min_{D_1,D_2,D_3,Z,Z_1,Z_2,M} \|X - D_1Z_1\|_F^2 + \lambda\|T - MZ\|_F^2 + \mu\left(\|Z_2 - \varphi(D_3Z)\|_F^2 + \|Z_1 - \varphi(D_2Z_2)\|_F^2\right) \quad (16)$$

Following the alternating direction method of multipliers we can segregate (16) into the following sub-problems. In ADMM, each of the sub-problems are for updating a single variable, the rest are assumed to be constant.

P1: $\min_{D_1} \|X - D_1Z_1\|_F^2$

P2: $\min_{D_2} \|Z_1 - \varphi(D_2Z_2)\|_F^2 \equiv \min_{D_2} \|\varphi^{-1}(Z_1) - D_2Z_2\|_F^2$

P3: $\min_{D_3} \|Z_2 - \varphi(D_3Z)\|_F^2 \equiv \min_{D_3} \|\varphi^{-1}(Z_2) - D_3Z\|_F^2$

P4: $\min_Z \lambda\|T - MZ\|_F^2 + \mu\|Z_2 - \varphi(D_3Z)\|_F^2$

$\equiv \min_Z \lambda\|T - MZ\|_F^2 + \mu\|\varphi^{-1}(Z_2) - D_3Z\|_F^2$

P5: $\min_{Z_1} \|X - D_1Z_1\|_F^2 + \mu\|Z_1 - \varphi(D_2Z_2)\|_F^2$

P6: $\min_{Z_2} \|Z_2 - \varphi(D_3Z)\|_F^2 + \|Z_1 - \varphi(D_2Z_2)\|_F^2$

$\equiv \min_{Z_2} \|Z_2 - \varphi(D_3Z)\|_F^2 + \|\varphi^{-1}(Z_1) - D_2Z_2\|_F^2$

P7: $\min_M \|T - MZ\|_F^2$

Note that the activation function is unitary (tanh), acts element-wise and hence is trivial to invert. That is the reason, we can express P2, P3 and P6 in their equivalent forms.

All the sub-problems are linear least squares problems in their original or their equivalent forms. They all have closed for solutions in the Moore-Penrose pseudo-inverse.

The problem is not convex, hence there is no global convergence guarantee. But this is the case for most machine learning models (with few exceptions like support vector machine). Here, we solve the sub-problems till a local minima is reached. We stop the iterations when the objective function does not change substantially with further iterations.

Algorithm for MLCDDL

Initialize: $D_1$, $D_2$ and $D_3$ randomly. From these, initialize $Z$ by solving $\min_Z \|X - D_1\varphi(D_2\varphi(D_3Z))\|_F^2$; the solution for the same is given in P8-P10. Initialize $Z_1 = \varphi(D_2\varphi(D_3Z))$ and $Z_2 = \varphi(D_3Z)$.

Iterate (k) till convergence –

Update: $M = TZ^T(Z^TZ)^{-1}$

Update: $D_1 = XZ_1^T(Z_1^TZ_1)^{-1}$

Update: $D_2 = \varphi^{-1}(Z_1)Z_2^T(Z_2^TZ_2)^{-1}$

Update: $D_3 = \varphi^{-1}(Z_2)Z_1^T(Z_1^TZ_1)^{-1}$

Update: $Z = (\lambda M^TM + \mu D_3^TD_3)^{-1}(\lambda M^TT + \mu D_3^T\varphi^{-1}(Z_2))$

Update: $Z_1 = (D_1^TD_1 + \mu I)^{-1}(D_1^TX + \varphi(D_2Z_2))$

Update: $Z_2 = (D_2^TD_2 + I)^{-1}(D_2^T\varphi^{-1}(Z_1) + \varphi(D_3Z))$

This concludes the algorithm for training. For testing, one needs to estimate the representation given the test sample x.

$$\min_z \|x - D_1\varphi(D_2\varphi(...\varphi(D_Nz)))\|_2^2 \quad (17)$$

This can be solved using substitutions as before. For first level it is $z_1 = \varphi(D_2\varphi(D_3z))$; for second level it is $z_2 = \varphi(D_3z)$. The augmented Lagrangian with these proxies will be expressed as,

$$\min_{z,z_1,...,z_N} \|x - D_1z_1\|_F^2 + \|z_1 - \varphi(D_2z_2)\|_F^2 + \|z_2 - \varphi(D_3z)\|_F^2 \quad (18)$$

Here the dictionaries learnt during the training process are used.

Using ADMM (18) is segregated into the following sub-problems.

P8: $\min_{z_1} \|x - D_1z_1\|_F^2 + \|z_1 - \varphi(D_2z_2)\|_F^2$

P9: $\min_{z_2} \|z_1 - \varphi(D_2z_2)\|_F^2 + \|z_2 - \varphi(D_3z_3)\|_F^2$

$\equiv \min_{z_2} \|\varphi^{-1}(z_1) - D_2z_2\|_F^2 + \|z_2 - \varphi(D_3z_3)\|_F^2$

P10: $\min_z \|z_2 - \varphi(D_3z)\|_F^2 \equiv \min_z \|\varphi^{-1}(z_2) - D_3z\|_F^2$

All the sub-problems have a closed form solution in the form of pseudo-inverse.

Once the representation is obtained, it is multiplied by the learnt linear $M$ to obtain $t=Mz$. Using an empirical threshold, positions of all elements in $t$ above the threshold are considered as active classes for $x$.

*B. Multi-Label Consistent Deep Transform Learning*

The concept of deep dictionary learning has developed in the past year. Deep transform learning is even newer. There is only a single paper on this topic by the authors [35] – that too a greedy suboptimal one. This is the first work that introduces an optimal approach to deep transform learning, that too a supervised one.

The main idea of deep transform learning is similar to that of deep dictionary learning. Instead of analyzing the data by a single level of transform (7), multiple levels of transforms are used to produce the final level of coefficients. It is similar to a feed forward neural network. This is expressed as,

$$T_3\varphi(T_2\varphi(T_1X)) = Z \qquad (19)$$

Here $T_1$ operates on the data $X$ to produce the first level of coefficients. $T_2$ analyzes the first level of coefficient to produce the second level. Finally $T_3$ operates the second level of coefficients to generate $Z$. In the only work on this topic, this has been solved using a greedy sub-optimal approach.

In this work, we extend deep transform learning to its supervised version with a multi-label consistency terms. We solve the problem in an optimal fashion using the variable splitting Augmented Lagrangian approach followed by alternating direction method of multipliers.

The complete formulation is as follows,

$$\min_{T_1,T_2,T_3,Z,M} \|T_3\varphi(T_2\varphi(T_1X)) - Z\|_F^2 + \lambda \|T - MZ\|_F^2$$
$$+ \varepsilon \sum_{i=1}^{3} \left( \|T_i\|_F^2 - \log\det T_i \right) \qquad (20)$$

We substitute $Z_1 = \varphi(T_1X)$ and $Z_2 = \varphi(T_2Z_1)$. This leads to the following AL.

$$\min_{T_1,T_2,T_3,Z,Z_1,Z_2,M} \|T_3Z_2 - Z\|_F^2 + \lambda \|T - MZ\|_F^2$$
$$+ \varepsilon \sum_{i=1}^{3} \left( \|T_i\|_F^2 - \log\det T_i \right) \qquad (21)$$
$$+ \mu \left( \|Z_2 - \varphi(T_2Z_1)\|_F^2 + \|Z_1 - \varphi(T_1X)\|_F^2 \right)$$

Using ADMM we can break it down into the following sub-problems.

S1: $\min_{T_1} \mu \|Z_1 - \varphi(T_1X)\|_F^2 + \varepsilon \|T_1\|_F^2 - \log\det T_1$
$\equiv \min_{T_1} \mu \|\varphi^{-1}(Z_1) - T_1X\|_F^2 + \varepsilon \|T_1\|_F^2 - \log\det T_1$

S2: $\min_{T_2} \mu \|Z_2 - \varphi(T_2Z_1)\|_F^2 + \varepsilon \|T_2\|_F^2 - \log\det T_2$
$\equiv \min_{T_2} \mu \|\varphi^{-1}(Z_2) - T_2Z_1\|_F^2 + \varepsilon \|T_2\|_F^2 - \log\det T_2$

S3: $\min_{T_3} \|T_3Z_2 - Z\|_F^2 + \varepsilon \|T_3\|_F^2 - \log\det T_3$

S4: $\min_{Z} \|T_3Z_2 - Z\|_F^2 + \lambda \|T - MZ\|_F^2$

S5: $\min_{Z_1} \mu \left( \|Z_2 - \varphi(T_2Z_1)\|_F^2 + \|Z_1 - \varphi(T_1X)\|_F^2 \right)$
$\equiv \min_{Z_1} \|\varphi^{-1}(Z_2) - T_2Z_1\|_F^2 + \|Z_1 - \varphi(T_1X)\|_F^2$

S6: $\min_{Z_2} \|T_3Z_2 - Z\|_F^2 + \mu \|Z_2 - \varphi(T_2Z_1)\|_F^2$

S7: $\min_{M} \|T - MZ\|_F^2$

As in the case of synthesis deep dictionary learning, we have segregated the complex problem into 7 simpler sub-problems which are just least squares problems in their original or equivalent form. Expressing S1, S2 and S5 in the equivalent forms are trivial since the activations functions are easy to invert.

We have used two stopping criteria for the iterations. The first one is a limit on the maximum number of iterations. The second one is local convergence of the objective function. The convergence of problems such as ours via alternating direction method of multipliers have been recently proven in [36].

Algorithm for MLCDTL

Initialize: $T_1$, $T_2$ and $T_3$ randomly. From these, initialize $Z$, $Z_1$ and $Z_2$ by by applying $\varphi(T_1X) = Z_1$, $T_2\varphi(T_1X) = Z_2$ and $T_3\varphi(T_2\varphi(T_1X)) = Z$.

Iterate (k) till convergence –

Update: $M = TZ^T(Z^TZ)^{-1}$

Update: $T_1$ by solving S1 – given in (12)
Update: $T_2$ by solving S2 – given in (12)
Update: $T_3$ by solving S3 – given in (12)

Update: $Z = (\lambda M^TM + I)^{-1}(T_3Z_2 + \lambda M^TT)$

Update: $Z_1 = (T_2^TT_2 + I)^{-1}(T_2^T\varphi^{-1}(Z_2^T) + \varphi(T_1X))$

Update: $Z_2 = (T_3^TT_3 + \mu I)^{-1}(T_3^TZ + \mu\varphi(T_2Z_1))$

This concludes the training phase. Generating the representation from the test sample is simple for deep transform learning. One simply needs to use,

$$z = T_3\varphi(T_2\varphi(T_1x)) \qquad (22)$$

Once the representation is generated, the inference (classification) is drawn in a manner similar to deep transform learning.

*C. Computational Complexity*

For deep dictionary learning, all the sub-problems require solving a least square problem having a pseudo-inverse. The upper bound for complexity of computing it is $O(n^w)$ where n is the size of the matrix and w < 2.37; but note that this is an upper bound and is conjectured as w=2. In fact, if the sub-problems are solved by something like conjugate gradient, the complexity is exactly $O(n^2)$.

For deep transform learning we need to solve two kinds of problems. The updates for the transform require computing SVD which has a complexity of $O(n^3)$ and the updates for coefficients are least square problems, whose complexity has already been discussed. Therefore the overall complexity of this procedure is $O(n^3)$.

However, in practice, the training complexity is hardly of any importance. What is of essence is the complexity during run-time. The run-time complexity of deep dictionary learning is also $O(n^w)$ since it requires pseudo-inverses, see P8-P10. Furthermore it is iterative. But the computational complexity of deep transform learning is only $O(n)$ since it is non-iterative and only requires matrix products.

## IV. EXPERIMENTAL RESULTS

For evaluating the performance of the proposed multi-label classification algorithm two datasets have been used: Reference Energy Disaggregation Dataset (REDD) [37] and Pecan Street dataset [38]. The REDD dataset consists of both aggregated and appliance level power data from six houses at 1Hz. However we do not make use of the appliance level consumption data; we only need its state. To emulate real world conditions the samples are averaged over a time period of 1 minute for our experiments. Four high power consuming devices used in our experiments are dishwasher, kitchen outlet, lighting and washer dryer

Pecan street dataset consists of one minute appliance level and building level electricity data from 240 houses. For experiments subsets of 28 houses have been used. For this work 4 most power consuming devices (site meter, air conditioner, electric furnace and sockets) are used for experiments. To prepare training and testing data, aggregated and sub-metered data are averaged over a time period of 1 minutes Each training sample contains power consumed by a particular device in one day while each testing sample contains total power consumed in one day in particular house. 80% of the houses are assigned to training set and 20% to the test set.

Metrics used in traditional single class classification problems cannot be used for multi-label classification problems. Prior studies in this area [3], [5] proposed using three measures: macro F1, Micro F1 and energy error.

F1 score is widely used in single label classification problems and is defined as:

$$F1(TP, FP, FN) = \frac{2*TP}{2*TP + FP + FN}$$

Where TP is True positive, FP is false positive and FN is false negative.

F1 macro and F1 micro are the measures derived from F1 score. These are label based evaluation measures which depend on the averaging method (macro or micro) used [3], [5]. F1 macro measure is computed by averaging the F1 scores for each label. Whereas, F1 micro is computed after summing true positives, false positives and false negatives across all labels.

$$F1_{micro} = F1\left(\sum_{i=1}^{N} TP_i, \sum_{i=1}^{N} FP_i, \sum_{i=1}^{N} FN_i\right)$$

$$F1_{macro} = \frac{1}{N}\sum_{i=1}^{N} F1(TP_i, FP_i, FN_i)$$

Here, $TP_i$, $FP_i$ and $FN_i$ denote the number of true positives, false positive and false negative for the label i. N is the number of labels in the dataset.

The aforesaid measures are from the perspective of information retrieval. For us, a more useful metric would be the predicted energy consumption. For that, the energy error has been defined in [5]. It is defined as:

$$error = \frac{\left|\left(\sum_{i=1}^{N} Average\_power_i\right) - \left(\sum_{i=1}^{N} Actual\_power_i\right)\right|}{\left(\sum_{i=1}^{N} Actual\_power_i\right)}$$

We have used two datasets and benchmarked our proposed technique with several state-of-the-art papers – prior studies in NILM [3, 5] used MLKNN [9] and RAKEL [7]; these techniques use thresholded wavelet coefficients as input features. We compare with MLKNN and RAKEL as benchmarks.

Comparison has also been done with AFAMAP [39]; however for this technique only average energy error is reported since it is not a classification method and hence F1 measures cannot be computed.

Finally we compare with the multi-label consistent stacked autoencoder (MLCSAE) technique proposed in [17]. This is the only known prior study on deep learning based multi-label classification. We use the three layer architecture proposed in [17]. The number of nodes are halved in every layer following the usual rule of thumb in such cases.

For our proposed techniques the raw data of an hour's duration (therefore samples are of size 60) is used as input. In deep learning there is no principled way to choose the number of layers or the number of nodes in each layer; it is dependent on the collective experience of the researchers. Usually going deeper helps, but with limited volume of training data, going deeper also results in over-fitting; there is a trade-off. For moderate sized problems, as ours, a three layer architecture usually yields the best results. Therefore we have used such an architecture for MLCDDL (multi label consistent deep dictionary learning) and MLCDTL (multi label consisted deep transform learning). The number of basis used are 120-80-50 for MLCDDL and 120-80-40 for MLCDTL.

For MLCDDL, we need to specify only a single parameter 'λ'. This parameter controls the relative importance of the feature learning cost and the label consistency cost. Since there is no reason to favor one over the other, we have kept it to be unity. MLCDTL requires specification of two parameters λ and ε. For the same reason as MLCDDL, we keep λ=1. The value of ε controls the relative importance of the data fidelity term and the prior on the learnt transforms. We found that our method is robust to any value of this parameter between 0.01 and 1. Both the algorithms require specifying hyper-parameter µ; for arbitrary problems this needs to be tuned. But here the hyper-parameter carries a specific meaning; it controls the relative importance of the different layers. Since there is no reason to favor one layer over another, we argue that keeping µ=1 is a sensible choice. For both the techniques we have used tanh activation function.

The experimental results are shown in Tables I and II. Since we want to showcase the change in results with layers we show it for one to four layers. For the fourth layer, the number of atoms have been halved from the third layer in each case.

TABLE I
PERFORMANCE EVALUATION ON REDD DATASET

| Method | Macro F1-measure | Micro F1-measure | Average energy error |
|---|---|---|---|
| MLCDTL (1 layer) | 0.6738 | 0.6884 | 0.0983 |
| MLCDTL (2 layers) | 0.6814 | 0.6906 | 0.0539 |
| MLCDTL (3 layers) | **0.6981** | **0.7001** | **0.0366** |
| MLCDTL (4 layers) | 0.6901 | 0.6923 | 0.0453 |
| MLCDDL (1 layer) | 0.6798 | 0.6846 | 0.0944 |
| MLCDDL (2 layers) | 0.6857 | 0.6905 | 0.0592 |
| MLCDDL (3 layers) | **0.7020** | **0.7046** | **0.0316** |
| MLCDDL (4 layers) | 0.6951 | 0.6964 | 0.0427 |
| MLKNN | 0.5931 | 0.6034 | 0.1067 |
| RAKEL | 0.5334 | 0.5749 | 0.9948 |
| AFAMAP | - | - | 0.2149 |
| MLCSAE | 0.6237 | 0.6301 | 0.1145 |

TABLE II
PERFORMANCE EVALUATION ON PECAN DATASET

| Method | Macro F1-measure | Micro F1-measure | Average energy error |
|---|---|---|---|
| MLCDTL (1 layer) | 0.7039 | 0.7049 | 0.0236 |
| MLCDTL (2 layers) | 0.7094 | 0.7101 | 0.0201 |
| MLCDTL (3 layers) | **0.7104** | **0.7104** | **0.0115** |
| MLCDTL (4 layers) | 0.7096 | 0.7098 | 0.0169 |
| MLCDDL (1 layer) | 0.7065 | 0.7033 | 0.0275 |
| MLCDDL (2 layers) | 0.7089 | 0.7062 | 0.0223 |
| MLCDDL (3 layers) | **0.7100** | **0.7099** | **0.0178** |
| MLCDDL (4 layers) | 0.7047 | 0.7026 | 0.0248 |
| MLKNN | 0.6227 | 0.6263 | 0.0989 |
| RAKEL | 0.6620 | 0.6663 | 0.9995 |
| AFAMAP | - | - | 0.2371 |
| MLCSAE | 0.6641 | 0.6703 | 0.0361 |

In Tables 1 and 2, the aggregate results over all the houses are shown. We observe that our proposed deep methods significantly outperforms all other shallow and deep techniques both in terms of F1-score as well as energy error. The other observation is that, once we go deeper, the results improve from layers 1 to 3; but when we go even deeper, the problem of over-fitting arises and the results deteriorate. Of the pre-existing shallow techniques, RAKEL, although performs decent in terms of F1-score, is very poor in terms of energy error; MLKNN yields more balanced results. The AFAMAP yields better results than RAKEL but is much worse than MLKNN. The other deep learning technique MLCSAE outperforms the shallow ones but is worse than ours.

TABLE III
APPLIANCE LEVEL EVALUATION ON REDD DATASET

| Device | MLCDTL | | MLCDDL | | MLKNN | | RAKEL | | AFAMAP | MLCSAE | |
|---|---|---|---|---|---|---|---|---|---|---|---|
| | Error | F1-score | Error | F1-score | Error | F1-score | Error | F1-score | Error | Error | F1-score |
| Dishwasher | **0.0086** | **0.5722** | 0.0179 | 0.5697 | 0.1250 | 0.4937 | 0.9964 | 0.3413 | 0.2726 | 0.0851 | 0.5124 |
| Kitchen outlet | **0.0556** | **0.5731** | 0.0492 | 0.5826 | 0.1647 | 0.5202 | 0.9952 | 0.4645 | 0.3249 | 0.1092 | 0.5253 |
| Lighting | **0.0841** | **0.7068** | 0.0099 | 0.6907 | 0.1105 | 0.6384 | 0.9943 | 0.5975 | 0.1953 | 0.1006 | 0.6591 |
| Washer dryer | **0.0082** | **0.5702** | 0.0929 | 0.5648 | 0.1743 | 0.4304 | 0.9964 | 0.4302 | 0.1362 | 0.1149 | 0.5011 |

TABLE IV
APPLIANCE LEVEL EVALUATION ON PECAN DATASET

| Device | MLCDTL | | MLCDDL | | MLKNN | | RAKEL | | AFAMAP | MLCSAE | |
|---|---|---|---|---|---|---|---|---|---|---|---|
| | Error | F1-score | Error | F1-score | Error | F1-score | Error | F1-score | Error | Error | F1-score |
| Site Meter | **0.0113** | **0.8404** | 0.0278 | 0.8306 | 0.0696 | 0.8096 | 0.9995 | 0.7072 | 0.1761 | 0.0432 | 0.8140 |
| Air Conditioner | **0.0125** | **0.5286** | 0.0188 | 0.5164 | 0.1381 | 0.5063 | 0.9995 | 0.5041 | 0.2263 | 0.0605 | 0.5062 |
| Electric Furnace | **0.0059** | **0.5195** | 0.0229 | 0.5092 | 0.0899 | 0.5005 | 0.9995 | 0.4568 | 0.1104 | 0.0519 | 0.5001 |
| Socket | **0.0149** | **0.5589** | 0.0333 | 0.5595 | 0.2696 | 0.5483 | 0.9995 | 0.5071 | 0.3172 | 0.0824 | 0.5413 |

In Tables 3 and 4, we show the performances at the appliance level. We have shown results only for layer 3 since it yields the best results. We see that the conclusions remain the same. We yield the best results in terms of all metrics. RAKEL yields by far the worst results. MLCSAE improves upon MLKNN in terms of energy error but not in terms of F1-score. AFAMAP is only better than RAKEL.

## V. CONCLUSION

In this work, we propose a new approach to non-intrusive load monitoring based on the multi-label classification framework. There are several studies on the subject, that have used off-the-shelf algorithms to the said problem. The contributions of this work are far more fundamental. We propose two deep learning based techniques to address the said problem. Note that the deep learning techniques have been developed in a bottom-up manner for this study; we are not modifying any deep learning algorithm to solve our problem.

Our work is based on the dictionary learning and transform learning based approaches. Unsupervised versions of deep dictionary learning have been proposed before; however this is the first work on supervised deep dictionary learning. It is a multi-label classification framework, but as a special case it can solve the standard single label multi-class problem.

The formulation based on transform learning is new. There is only a single study on greedy sub-optimal deep transform learning for unsupervised problems. This is the first work that proposes an optimal supervised version. As in dictionary learning, the single label multi-class problem is a special case of our proposed work.

In this work, we posed non-intrusive load monitoring as a multi-label classification problem. It is based on the assumption that during training only the aggregate data and the logs (ON/OFF state) of the different devices are available. This assumption was made in order to depict less intrusive training scenarios. However, if we assume that the complete data, i.e. device levels power consumptions are available – instead of classification we can pose the problem as a multi-variate regression problems. Our proposed frameworks can naturally handle it; instead of having 1 / 0 labels in the targets we will have the corresponding power consumption levels for each device. During testing, the learnt model would directly predict the power instead of the state. Experimental results on real datasets show that our proposed method surpasses all popular and state-of-the-art techniques in multi-label classification.

In this work, we have assumed that the appliance level power consumption is not known, only their state is given. But if we consider the appliance level consumption during training, we can formulate NILM as a multi-variate regression problem. In that case, instead of predicting the state (which is a binary value), we will be predicting power consumption (real positive value). The methodology developed in this work can automatically handle the said scenario. In the targets, instead of recording the states, we need to have the power consumptions; the rest would remain the same. We would like to try this approach in the future.

ACKNOWLEDGEMENT

This work is partially supported by the Infosys Center for Artifical Intelligence at IIIT Delhi and by the DST IC-IMPACTS Indo-Canadian Grant.